# Quantifying the Strategyproofness of Mechanisms
## via Metrics on Payoff Distributions

**Benjamin Lubin**
Harvard University, SEAS
Cambridge, MA 02138
blubin@eecs.harvard.edu

**David C. Parkes**
Harvard University, SEAS
Cambridge, MA 02138
parkes@eecs.harvard.edu

## Abstract

Strategyproof mechanisms provide robust equilibrium with minimal assumptions about knowledge and rationality but can be unachievable in combination with other desirable properties such as budget-balance, stability against deviations by coalitions, and computational tractability. In the search for maximally-strategyproof mechanisms that simultaneously satisfy other desirable properties, we introduce a new metric to quantify the strategyproofness of a mechanism, based on comparing the payoff distribution, given truthful reports, against that of a strategyproof "reference" mechanism that solves a problem relaxation. Focusing on combinatorial exchanges, we demonstrate that the metric is informative about the eventual equilibrium, where simple regret-based metrics are not, and can be used for online selection of an effective mechanism.

## 1 Introduction

Mechanism design addresses the problem of achieving desirable outcomes in multi-agent systems despite private information about valuations and individual self-interest. Mechanism design finds applications in societal contexts (e.g., school and medical residents matching [1]) and business contexts (e.g., sponsored search auctions [9]), while providing a formal paradigm by which to coordinate the behavior of artificial agents (e.g., for task and resource allocation). A central concept is that of *strategyproofness*: is there a desirable mechanism in which it is a dominant-strategy equilibrium for every agent to report its private information (or *type*) truthfully? Strategyproofness simplifies participation and removes the need for counterspeculation about the behavior of other agents. But strategyproofness can be unachievable together with other desirable properties. For example, strategyproofness can conflict with other desired properties such as budget-balance [14], coalitional stability or revenue properties [3], simple rules [9], and computational tractability [13]. In addition, there are some problems for which the design of a strategyproof mechanism with desirable properties is unattainable with current theoretical techniques. For these reasons, it is often necessary to adopt approximately strategyproof mechanisms, and for this it is useful to have a metric to quantify the degree of strategyproofness of a mechanism to guide the design process.

A standard measure of approximate strategyproofness is *regret*, namely the loss in utility to an agent from reporting its true type compared to its best possible misreport, given reports of other agents. An *$\epsilon$-strategyproof mechanism* is one in which truthful reporting achieves within $\epsilon > 0$ of the best possible utility, for all possible reports of other agents and all agent types [17]. This is meaningful when $\epsilon$ is small, for example smaller than the cost an agent incurs in reasoning about how to manipulate, because it is reasonable that agents will then behave truthfully. But, as the maximal regret gets large it is not clear that regret provides the appropriate metric by which to quantify the degree of strategyproofness of a mechanism or guide mechanism design.

Conceptually, one could imagine simply defining a metric on the distance between equilibrium strategies and truthful strategies. But this metric provides no guidance for how to design approximately strategyproof mechanisms. Moreover, we are interested in a metric that does not require solving for the equilibrium of a candidate mechanism, because this will tend to be the bottleneck in computational approaches to identifying good mechanisms. We introduce as a metric the *normalized Kullback-Lieber (KL) distance* between the distribution of payoffs in a mechanism and a distribution induced by a strategyproof "reference" mechanism, where these payoffs (or utilities) are evaluated given truthful bids (i.e. out of equilibrium), and restricted to agents affected by the outcome (either positively or negatively.) The metric requires that there exists a strategyproof reference mechanism for some natural relaxation of the problem.

In studying this KL-distance metric, we focus on mechanism design for combinatorial exchanges (CEs), which extend combinatorial auctions to allow for multiple buyers and multiple sellers. The design of highly efficient, strategyproof CEs remains an open problem in mechanism design.[1] The Vickrey-Clarke-Groves mechanism is strategyproof and efficient, but runs at a deficit. This provides the reference mechanism. We evaluate the KL-distance metric and a number of regret-based metrics of a family of approximately strategyproof mechanisms that were proposed in Parkes et al. [14]. In providing experimental results, we need to adopt an approximate method to compute equilibrium of different CE mechanisms because there is no computationally tractable method to compute exact Bayesian-Nash equilibrium in CEs. For this, we compute restricted, partially-symmetric equilibria.

The KL-distance metric has a significant and strongly positive correlation with a parametrization of the amount by which the equilibrium deviates from truthful reports, and a strongly negative correlation with the allocative efficiency in equilibrium. The metric identifies the Small rule from Parkes et al. [14] as the best mechanism, and it is indeed this rule that provides highest efficiency and least bid-shaving in equilibrium. In testing the power of the metric for mechanism design, we show that the metric is effective in guiding a search through a set of mechanisms and identifying a highly efficient mechanism based only on observed data. In closing, we discuss the implications of the metric for advancing a new paradigm of heuristic mechanism design and also present a number of open questions.

**Related Work.** Schummer [17] was the first to consider $\epsilon$-strategyproof mechanisms and this approach was also considered by Kothari et al. [8] in the design of multi-unit auctions. In doing so, these authors advocate *worst-case regret* as a metric of approximate strategyproofness, namely the worst-case loss in utility from behaving truthfully given all possible reports of other agents. Another notion of approximate strategyproofness is that of *strategyproof with high probability* [2]. The aforementioned body of work is generally motivated by problems in which the approximation can be arbitrarily small. Alternatively, Parkes et al. [14] first advocated the idea of defining a payment rule that tries to minimize the distance to the payments in the VCG mechanism, in settings such as CEs for which the VCG payments are unavailable because they run at a deficit. This approach has been adopted and expanded upon in the context of combinatorial auctions, where core constraints can often preclude VCG payments [11].[2] Our work also relates to methods of *automated mechanism design* [6], in which systematic search is performed in the space of possible mechanism rules, and especially *empirical mechanism design* [18, 19], in which one couples search through a parametrized mechanism space with an empirical methodology for solving the induced games.

## 2 A Heuristic Mechanism Design Paradigm

In the problem of mechanism design, there is a set of alternatives $A$ and a set of agents $N = \{1, \ldots, n\}$ and each agent has a private valuation function $v_i(a) \in \mathbb{R}$ for each alternative. In the context of this paper, each alternative represents a trade of goods between agents. We consider here the standard setting of quasi-linear utility functions, where an agent's utility (or *payoff*) for alternative $a$ and payment $p$ is $u_i(a, p) = v_i(a) - p$. A direct-revelation mechanism asks each agent to make a claim about its valuation, from which an alternative $f(\hat{v}) \in A$ is picked based on claims $\hat{v} = (\hat{v}_1, \ldots, \hat{v}_n)$ on valuations, and payments $p_i(\hat{v}) \in \mathbb{R}$ are collected from each agent. A *strategyproof* mechanism is one in which it is a dominant-strategy for each agent to report its true valuation, so that $v_i(f(v_i, v_{-i})) - p_i(v_i, v_{-i}) \geq v_i(f(\hat{v}_i, v_{-i})) - p_i(\hat{v}_i, v_{-i})$, for all $v_i$, all $\hat{v}_i$, and all $v_{-i} = (v_1, \ldots, v_{i-1}, v_{i+1}, \ldots, v_n)$.

In motivating the need for a metric to quantify approximate strategyproofness, consider the following heuristic approach to mechanism design: there is a space of non-strategyproof mechanisms $\mathcal{M}$, each of which has the same outcome rule and good properties when agents are truthful, and with properties that degrade as agents becomes less truthful in equilibrium. *Given this set of mechanisms, adopt as the goal that of selecting the mechanism in $\mathcal{M}$ that is maximally strategyproof.* For example, these could be mechanisms in which outcome rule $f(v) \in \arg\max_{a \in A} \sum_i v_i(a)$ but vary in their payment rules, so that if agents are truthful the mechanism is efficient; i.e., maximizing the total value through its choice of alternative. In doing so, we seek a metric on approximate strategyproofness that provides explicit design guidance because the space of mechanisms may be too large to enumerate, and works without computing the equilibrium of a candidate mechanism because this is computationally expensive.

A standard answer would be to select a mechanism that minimizes the worst-case *ex post* regret from behaving truthfully, across all agents and across all instances. The regret of agent $i$ when valuations are $v = (v_1, \ldots, v_n)$ is $regret_i(v) = \max_{\hat{v}_i}(v_i(f(\hat{v}_i, v_{-i})) - p_i(\hat{v}_i, v_{-i})) - (v_i(f(v_i, v_{-i})) - p_i(v_i, v_{-i}))$. But is this the right answer? Does this lead to a mechanism in which an agent's equilibrium bids are closer to truthful, on average, than in the other

---

[1]It is well known that no mechanism exists that is efficient, no deficit and individual rational [12]. But no "second best" mechanism has been designed that maximizes expected efficiency while retaining incentive compatibility, individual-rationality and no deficit properties.

[2]Budish [4] recently advocated "strategyproofness in a large-market" as a criteria for selecting amongst two, non-strategyproof mechanisms. This asks whether the mechanism will become strategyproof for a replica economy, in the limit as each agent be-

comes one of a continuum of agents with the same type. While a very useful design criteria, this does not by itself meet our needs of providing a metric with which to quantify approximate strategyproofness.

mechanisms in $\mathcal{M}$? In this paper, we propose a metric that adopts a strategyproof *reference mechanism* $m^*$, and seeks a mechanism that induces payoffs that are close in distribution to $m^*$. The reference mechanism will be outside of $\mathcal{M}$, and with the same outcome rule but a payment rule that makes the mechanism strategyproof.

## 3 The Metric and the CE Environment

The metric is defined as a KL-distance between payoff distributions to agents in a mechanism $m = (f, p)$ and its reference, strategyproof mechanism $m^*$. For a particular instance, let $\pi^m(v) = (\pi_1(v), \ldots, \pi_n(v))$ define the payoff to each agent in $m$, i.e. $\pi_i(v) = v_i(f(v)) - p_i(v)$. Similarly, let $\pi^*(v) = (\pi_1^*(v), \ldots, \pi_n^*(v))$ define the payoff to each agent in the reference mechanism $m^*$. Let $\pi \in \Pi$ be a feasible joint payoff vector and let $H^m(\pi)$, $H^*(\pi)$ be the joint distribution of payoffs under mechanism $m$ and $m^*$ respectively, as induced by a distribution on valuations. In general, we have in mind a metric defined as the multivariate KL-distance between these distributions: $\int_{\pi \in \Pi} H^*(\pi) \log(\frac{H^*(\pi)}{H^m(\pi)}) d\pi$. To keep things relatively simple, we will consider in this paper a projection of these multi-dimensional distributions down to one-dimensional, normalized payoff distributions where the normalization is based on a relevant statistic for a particular instance. The particular projection is specific to the CE environment.

### 3.1 Combinatorial Exchanges

A CE is a market with multiple units of dissimilar, indivisible items, $G = \{1, \ldots, k\}$, and multiple agents, each of which may be interested in both buying and selling items. Each agent $i$ has a valuation $v_i(\lambda_i) \in \mathbb{R}$ on possible trades $\lambda_i = (\lambda_{i1}, \ldots, \lambda_{ik})$, where $\lambda_{ij} \in \mathbb{Z}$ specifies the number of units of item $j$ transferred to agent $i$. An efficient CE will identify the trade that maximizes the total value across all feasible trades, subject to feasibility constraints (supply $\geq$ demand). The Vickrey-Clarke-Groves (VCG) mechanism adopts the role of the reference mechanism by relaxing the no-deficit constraint. Given reported valuations $\hat{v}$, the VCG selects the efficient trade $\lambda^*$ based on reports, to maximize the total value over all feasible trades (this problem can be formulated and solved as a mixed-integer program). Let $V^*(\hat{v})$ denote the total value (or *surplus*) over all agents in this trade. In the VCG mechanism, each agent's payment is $p_{\text{vcg},i}(\hat{v}) = \hat{v}_i(\lambda^*) - (V^*(\hat{v}) - V^*(\hat{v}_{-i}))$, where $V^*(\hat{v}_{-i})$ is the total reported value for the optimal trade without the presence of agent $i$. The VCG mechanism is strategyproof, but runs at a deficit.

Recognizing this, Parkes et al. [14] introduced a number of approximately SP mechanisms, defined for CEs. These will play the role of the design space $\mathcal{M}$ in this paper. Each mechanism adopts the same allocation rule as in VCG (and therefore has good properties when agents are truthful) but defines payments that are exactly bal-anced. Conceptually, the payment rules all discount the amount an agent $i$ will pay relative to its reported valuation $\hat{v}_i(\lambda^*)$ for the selected trade. In the VCG mechanism, this discount is $\Delta_{\text{vcg},i}(\hat{v}) = V^*(\hat{v}) - V^*(\hat{v}_{-i})$, but in each of these new mechanisms the discounts are constrained so that $\sum_i \Delta_i(\hat{v}) = V^*(\hat{v})$, providing $\sum_i p_i(\hat{v}) = \sum_i (\hat{v}_i(\lambda^*) - \Delta_i(\hat{v})) = V^*(\hat{v}) - V^*(\hat{v}) = 0$ and no-deficit. The deviation from the payments of the VCG mechanism opens up the possibility that an agent can gain by deviating from its truthful report. The regret of agent $i$ is exactly $regret_i(\hat{v}) = \Delta_{\text{vcg},i} - \Delta_i(\hat{v})$, i.e. the amount by which the discount is less than that in the VCG mechanism.

Each mechanism in $\mathcal{M}$ adopts a different method to allocate the available surplus to agents. The mechanisms that we consider are: *Two Triangle, Threshold, Reverse, Large, Small, Fractional,* and *Equal*. The details are presented in the Appendix. For now, we simply note that the Threshold rule has been considered of particular interest because it defines payments that minimize the maximal regret to agents, given the no-deficit constraint. Connecting back to the earlier notation, we can also observe that the payoff $\pi_i(v)$ to agent $i$ in instance $v$, and when agents are truthful, is simply its discount $\Delta_i(v)$ while the payoff in the reference mechanism is $\Delta_{\text{vcg},i}(v)$.

### 3.2 The KL-Distance Metric and Other Metrics

In the CE environment, we specialize the general multivariate KL-distance to a KL-distance on *normalized* payoff, where the payoff $\pi_i^m(v)$ to each agent in instance $v$ is normalized by $V^*(v)$, the total available surplus that constrains the total available discounts provided to agents. Given this, the normalized KL-distance metric for mechanism $m$ is defined as:

$$KLnorm(m) = \int_0^\infty \widehat{H}^*(\pi) log \left( \frac{\widehat{H}^*(\pi)}{\widehat{H}^m(\pi)} \right) d\pi, \quad (1)$$

where $\widehat{H}^*(\pi)$ is the univariate distribution of the normalized payoff $\frac{\pi_i^*(v)}{V^*(v)}$ under the reference mechanism, given the distribution on instances, and $\widehat{H}^m(\pi)$ is similarly defined for the mechanism being considered. We further restrict these distributions to payoffs associated with agents that are active in the efficient trade. Note that the distribution on payoffs is that induced by the *true* distribution on valuations, not by the equilibrium distribution. We also consider an unnormalized KL-distance metric.

In addition, we adopt a number of regret-based metrics:

$$L_1(m) = \int_v ||\pi_+^*(v), \pi_+^m(v)||_1 \, g(v) dv \quad (2)$$

$$L_1 norm(m) = \int_v ||\frac{\pi_+^*(v)}{V^*(v)}, \frac{\pi_+^m(v)}{V^*(v)}||_1 \, g(v) dv \quad (3)$$

$$L_2(m) = \int_v ||\pi_+^*(v), \pi_+^m(v)||_2 \, g(v) dv \quad (4)$$

$$L_2 norm(m) = \int_v || \frac{\pi_+^*(v)}{V^*(v)}, \frac{\pi_+^m(v)}{V^*(v)}||_2 \, g(v)dv \quad (5)$$

$$L_\infty(m) = \int_v ||\pi_+^*(v), \pi_+^m(v)||_\infty \, g(v)dv \quad (6)$$

$$L_\infty norm(m) = \int_v || \frac{\pi_+^*(v)}{V^*(v)}, \frac{\pi_+^m(v)}{V^*(v)}||_\infty \, g(v)dv \quad (7)$$

where $g(v)$ is the p.d.f. on valuation instances $v$ (for the truthful distribution), $\pi_+^*(v)$ and $\pi_+^m(v)$ indicate the payoff vectors restricted to agents that are active in the trade, and $L_1(\cdot, \cdot), L_2(\cdot, \cdot), L_\infty(\cdot, \cdot)$ are standard $L_1, L_2$ and $L_\infty$ metrics. Note that although all of the metrics above are defined over a continuous valuation space, practical evaluation will require numerical integration over samples.

### 3.3 An Initial Evaluation in Three CE Scenarios

We consider three CE generators, and thus three different problem scenarios.[3] Two are variations on the combinatorial auction generators (*Decay* and *Uniform*) introduced in Sandholm [16]. To make these work in an exchange setting, we first fix the set of available goods and then distribute them to the selling agents, and the demand for them among the buying agents. With these endowments and 'demand sets' specified, we then choose negative seller (reserve) values, and positive buyer values for XOR bundles of items restricted to these endowments and 'demand sets', according to Sandholm's rules. The third is a new generator (*Super*), specifically designed for CEs, and with features carefully crafted for super-additive valuations. Here every good $g \in G$ is assigned a uniform random common value $c(g) \geq 0$, and a uniform random private value specific to agent $i$, $y_i(g) \geq 0$. Agent $i$ then has a value for an individual good $w_i(g) = \beta y_i(g) + (1 - \beta)c(g)$, for some $\beta$ (.5 in our experiments). The value to agent $i$ for all bundles of items $S \subseteq G_i$ is then $(\sum_{g \in S} w_i(g))^\gamma$, for some $\gamma > 1$, where $G_i$ is the endowment/'demand set' for agent $i$. As above, this value forms a negative (reserve) value for sellers and a positive value for buyers.[4]

It is instructive to consider the distribution of $V^*(v)$, $V^*(v_{-i})$, and the VCG payoff $V^*(v) - V^*(v_{-i})$ for trading agents that is induced by these generators. See Figures 1 and 2 for the *Super* distribution (the others are qualitatively similar). We can precisely identify the form of these distributions. Fix instance $v$. Consider the set $\Lambda$ of feasible trades in a given market instance. Each $\lambda \in \Lambda$ has a corresponding total value $V(\lambda, v)$, and $V^*(v)$ is by definition the maximum over these. Thus the $V^*$ distribu-tion is that of the extreme values of the underlying distribution of $V$. Such extreme value distributions have been extensively studied in the statistics literature, and can be precisely modeled by the *Generalized Extreme Value Distribution* (GEV) p.d.f. Figure 1 shows the excellent fit of the GEV that can be produced for both $V^*$ and $V_{-i}^*$ via maximum likelihood estimation (MLE). The VCG payoff distribution is the distribution of exceedences (by $V^*$) over $V_{-i}^*$, and is well-modeled by a *Generalized Pareto Distribution* (GPD), though this model is typically motivated in cases of exceedences over a fixed threshold. The MLE fit of the GPD is illustrated in Figure 2, along with the fit of a simple Exponential distribution (which is generalized by the GPD), indicating that the extra parameters of the GPD are improving the fit.

---

[3] Please contact the authors to obtain our data sets and specific parametrizations.

[4] We do not use CATS [10] for the generation of our data sets because its algorithms are explicitly designed for auctions and it is not straightforward to extend its distributions in a way that appropriately balances buyers and sellers. In the absence of such reference distributions, we have opted for these simpler existing generators, coupled with our own new generator.

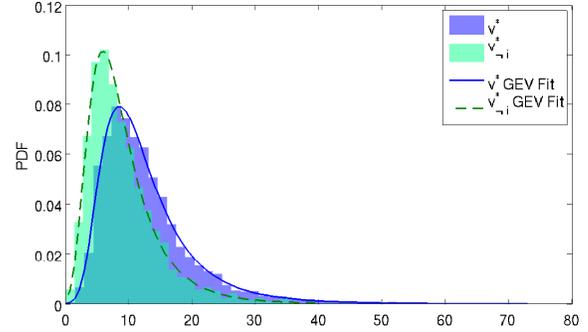

Figure 1: Distribution of surplus and marginal-surplus

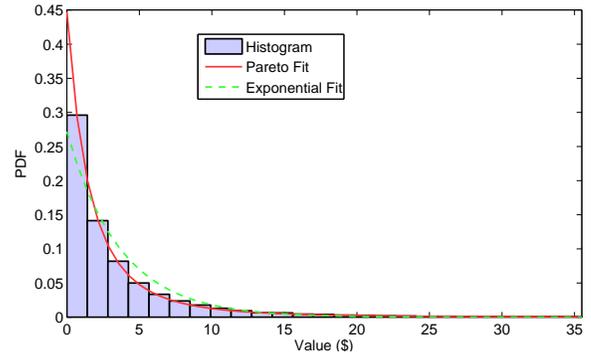

Figure 2: Distribution of VCG payoffs

We can immediately consider how well each of the mechanisms performs at mimicking this distribution of payoffs. Figure 3 shows an empirical c.d.f. of the payoff to trading agents under each of mechanism, when agents behave truthfully (again for the *Super* generator, the others being similar). One can visually confirm that the Small rule is the one best tracking the VCG payoffs in distribution. Table 1 computed the normalized metrics on each mechanism, computed over all three scenarios. Consistent with Figure 3, we can observe that Small has the smallest $KLnorm$ metric. On the other hand, Threshold has the smallest $L_2 norm$ and $L_\infty norm$ (regret-based) metrics. Notice that the $L_1 norm$ metric is identical across all rules except, No Discount and Equal. This is because the other mechanisms always allocate all available surplus as payoff to agents.[5]

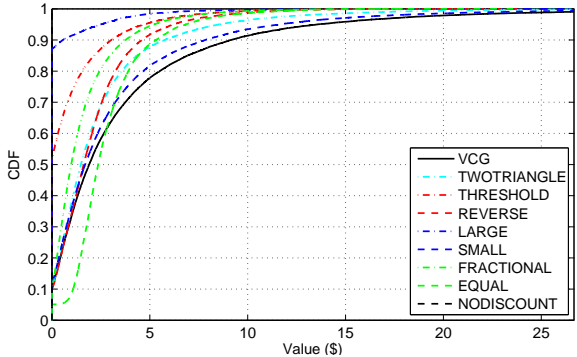

Figure 3: Distribution of payoffs in each mechanism

| Mechanism | $KL\,norm$ | $L_1\,norm$ | $L_2\,norm$ | $L_\infty\,norm$ |
|---|---|---|---|---|
| Two Triangle | 0.0735 | **0.5914** | 0.3170 | 0.1917 |
| Threshold | 0.0472 | **0.5914** | **0.2355** | **0.1016** |
| Reverse | 0.1251 | **0.5914** | 0.3066 | 0.2210 |
| Small | **0.0452** | **0.5914** | 0.4208 | 0.3527 |
| Large | 0.0559 | **0.5914** | 0.3110 | 0.2070 |
| Fractional | 0.0741 | **0.5914** | 0.2528 | 0.1513 |
| Equal | 0.3043 | 0.8037 | 0.3727 | 0.2576 |
| No Discount | 0.6372 | 1.5876 | 0.6679 | 0.4030 |

Table 1: Metric value at truth averaged across all three CE scenarios. Minimal metric values in **bold**.

## 4 Equilibrium Analysis

Computing the equilibrium of the various mechanisms presents a challenge because this is an infinite game of incomplete information, with a continuum of possible valuations and thus possible agent strategies. The game also has combinatorial structure. There are at present no tractable methods to compute the exact Bayes-Nash equilibrium for such problems. The state of the art approach is to search for parametrized strategy profiles that constitute a restricted equilibrium through iterated best-response dynamics [20].[6] This is the approach that we adopt here, with adaptive grid search to compute a best-response.

### 4.1 Computing Restricted Bayes-Nash Equilibrium

One simple restriction that one could impose is that every agent shaves its valuation by $\alpha \geq 0$, and thus seek a symmetric Bayes-Nash equilibrium. In the context of a CE, agents would report valuations $(1 - \alpha)v$ and $(1 + \alpha)v$ for buyers and sellers respectively (note that sellers

---

have negative values.) This simplification realizes a one-dimensional, continuous strategy space.

We compute a more fine-grained equilibrium by also running experiments in which we adopt two or three shave factors. With multiple shave factors, we associate each agent in an instance endogenously with a valuation class depending on its valuation function. For example, with three shave factors $\alpha_1, \alpha_2$, and $\alpha_3$, we sort agent valuations into "low," "medium" and "high" valuation classes, with an agent in each class associated with shave factor $\alpha_1, \alpha_2$ and $\alpha_3$ respectively. We then search for an equilibrium defined in terms of these three parameters. To sort agent valuations, we first draw a number of samples of otherwise unused agents from the same distribution that defines the CE scenario, and for each of these agents, we record the 95th percentile of value across the trades that define its valuation function. An agent's valuation class is identified by comparing the value at the 95th percentile on the trades in its valuation with the sampled values, and assigning a class according to placement in the lower, middle, or upper third (tritile) of this sampled distribution.

For any number of shave factors, our algorithm for finding the equilibrium begins with provisional shave factors $\{\widehat{\alpha}_k\}$ (e.g., for $k \in \{1, 2, 3\}$) set to 0. It then repeatedly generates a set of CE instances from the particular distribution (Uniform, Decay or Super), and for each instance, each agent is first placed into a valuation class when using multiple shave factors. In each iteration $t$ of the algorithm, and for each agent $i$, a grid search is performed on $\alpha$-values to find its best-response value $\widetilde{\alpha}_i$, while using provisional $\alpha$-values assigned to the other agents. For each valuation class, the provisional $\widehat{\alpha}_k$ are then updated as $\widehat{\alpha}_k^{t+1} := \theta \widehat{\alpha}_k^t + (1 - \theta) \overline{\alpha}_k^t$, where $\theta = .5$ and $\overline{\alpha}_k^t$ is the mean of the best response values in iteration $t$ calculated for each agent associated with the class $k$. The width of the grid search in period $t + 1$ is chosen endogenously, with 10 points covering a span of $|\widehat{\alpha}_k^t - \overline{\alpha}_k^t|$. Search stops when this error estimation falls below a fixed constant $\kappa = 0.001$.

### 4.2 Equilibrium: Results

Table 2 shows the results with one-dimensional and three-dimensional strategy spaces (respectively "one class" and "three classes"), for all three generators. In the case of three classes, the reported shave factor is the average across $\{\alpha_1, \alpha_2, \alpha_3\}$. The best mechanisms in each case are indicated in **bold**. Surprisingly, the Threshold mechanism, which has some theoretical support in minimizing the *ex post* regret across all these mechanisms, does not perform nearly as well as the Small mechanism either in terms of the size of shave factor (close to zero indicates approximate incentive-compatibility) or the resulting allocative efficiency. Recall that the Small mechanism is also the one

|  | One Equilibrium Class | | | | | | Three Equilibrium Classes | | | | | |
|  | Shave Factor | | | Efficiency (%) | | | Shave Factor | | | Efficiency (%) | | |
| **Rule** | **Dec.** | **Uni.** | **Sup.** | **Dec.** | **Uni.** | **Sup.** | **Dec.** | **Uni.** | **Sup.** | **Dec.** | **Uni.** | **Sup.** |
| *VCG* | *0.0* | *0.0* | *0.0* | *100* | *100* | *100* | *0.0* | *0.0* | *0.0* | *100* | *100* | *100* |
| Two Triangle | **0.1** | **0.2** | **0.6** | **99.99** | **100** | **99.99** | **0.1** | **0.4** | **5.6** | **99.99** | **100** | 97.95 |
| Threshold | 12.0 | 28.7 | 10.7 | 99.09 | 97.43 | 98.01 | 14.6 | 27.2 | 11.2 | 93.64 | 81.09 | 89.74 |
| Reverse | 14.9 | 57.7 | 52.3 | 98.70 | 83.38 | 51.52 | 13.0 | 65.8 | 57.6 | 98.99 | 77.30 | 56.08 |
| Small | **0.1** | **0.2** | **0.3** | **99.99** | **100** | **100** | **0.0** | **0.1** | **0.2** | **99.99** | **100** | **100** |
| Large | **2.6** | **2.3** | 9.8 | **99.96** | **99.99** | 98.26 | **2.8** | **2.9** | 67.1 | **99.96** | **99.98** | 78.83 |
| Fractional | 71.2 | 71.1 | 53.0 | 59.39 | 67.34 | 49.07 | 62.7 | 81.9 | 62.0 | 37.12 | 63.09 | 56.77 |
| Equal | 75.4 | 77.6 | 52.5 | 51.96 | 55.76 | 51.01 | 62.2 | 78.3 | 66.8 | 33.35 | 54.21 | 52.19 |
| No Discount | 75.6 | 76.0 | 53.2 | 51.56 | 59.01 | 48.23 | 62.3 | 80.9 | 72.4 | 34.15 | 50.11 | 48.21 |

Table 2: Restricted Bayes-Nash equilibrium: Shave Factor and Allocative Efficiency in Each Mechanism.

with the lowest KL-distance metric.[7]

To understand the effect of the Small payment rule, which allocates payment preferentially to agents with a small VCG payoff, we can study an individual agent's incentive to deviate. Figure 4 shows the profit gained by a single agent in a representative single instance drawn from the *Super* scenario, as the agent reports $V_R$ compared to truth $V_T$ for its winning trade and 0 for all other trades, under each of the mechanisms. The profit is normalized to its maximal possible profit, i.e. its VCG profit, and the experiment considers only unilateral deviation by this agent with all other agents reporting truthfully. The agent in question has a large payoff under VCG, which the Large mechanism fully allocates. As the agent deviates he suffers a loss under the Large mechanism. Under all the other mechanisms (except VCG) there is at least some gain from deviation. Unlike the other rules, though, the Small mechanism exhibits a flat plateau once the agent deviates by a small amount. Thus the incentives to deviate significantly can be quite low under Small, even for agents whose payoff in VCG is quite large.

This analysis represents only a single agent in a single instance. In order to get a more comprehensive picture we can average several thousand such single-instance trajectories, as shown in Figure 5. Here we see that mis-reporting makes an agent strictly worse-off under VCG, as expected. But importantly, we see that the Small mechanism provides only a small expected gain from deviation, and the maximal expected gain occurs with less shaving then the other mechanisms. While still a non-equilibrium analysis (other agents are truthful), this is suggestive of the good equilibrium performance under Small.

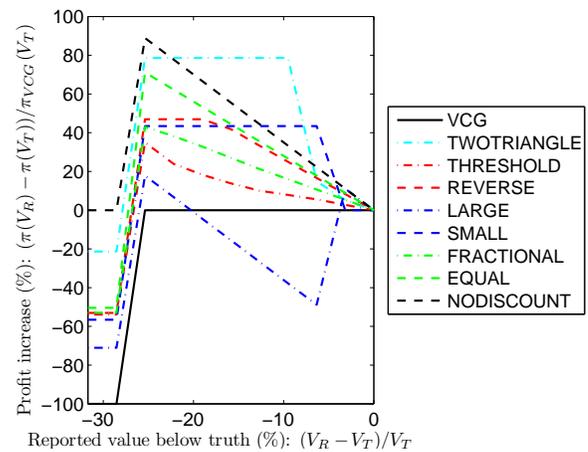

Figure 4: Profit gain by unilateral mis-report.

In determining a good strategy, an agent is in essence making an *ex ante* trade-off between potential gain from a successful manipulation and potential loss given an unsuccessful manipulation. By further conditioning on those misreports that are successful (i.e., when an agent still trades) and unsuccessful, we arrive at Figures 6 and 7. We see that Small is near the bottom of the pack for both conditional gain and conditional loss, indicating that success brings relatively less gain while failure brings relatively more pain than in other mechanisms. In comparison, an unsuccessful manipulation does not hurt an agent as much under the Threshold mechanism, contributing to its weaker equilibrium performance.

**Remark.** Unlike Small, the Threshold mechanism tends to allocate payoff to fewer agents, and with very few (if any) agents receiving their maximal payoff. This is driving the divergence from the VCG payoff distribution and also this larger loss in payoff, conditioned on an unsuccessful manipulation. By making the distribution on payoffs

---

[7]One interesting anomaly in the data is for Large between the "one class" and "three class" analysis. With one class, a balance must be made in the equilibrium between those agents with high valuations (likely to receive their full discount without any shave under Large) vs. those with low valuations (unlikely to receive any discount without shaving). In this case, the former constrains the latter and agents choose not to shave much in equilibrium. But with three shave factors there is increased discrimination, and the optimal shave for those with small valuations becomes very extreme. This, coupled with the fact that there are large numbers of small discounts relative to a few large discounts, decreases the efficiency of the Large rule in equilibrium.

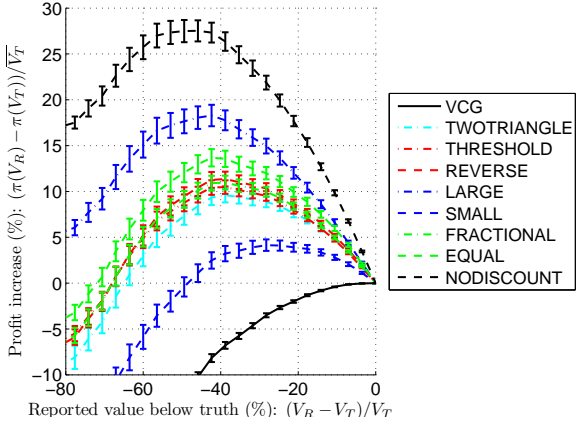

Figure 5: Expected profit by unilateral mis-report

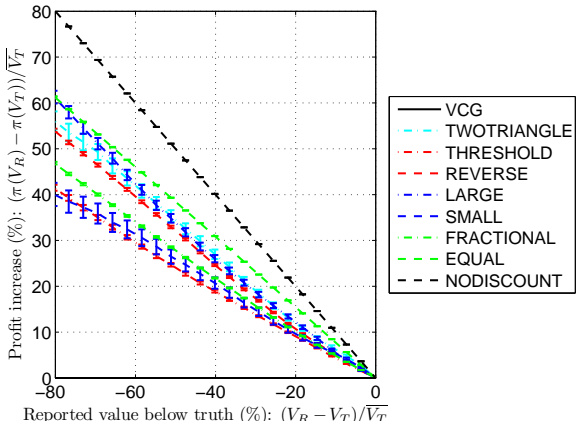

Figure 6: Conditional profit by unilateral mis-report

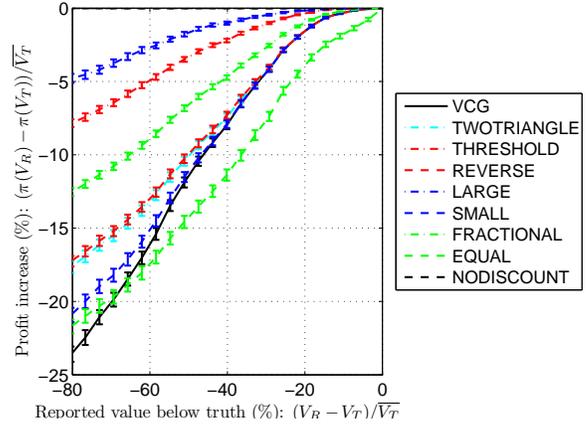

Figure 7: Conditional loss by unilateral mis-report

| Correlation with Efficiency at Truth | | | |
|---|---|---|---|
| **Metric** | **Corr.** | **$\rho$-value** | **Significant?** |
| $KLnorm$ | -0.3814 | 0.0044 | Y |
| $L_1\,norm$ | -0.1698 | 0.2197 | N |
| $L_2\,norm$ | 0.0154 | 0.9120 | N |
| $L_\infty\,norm$ | 0.0220 | 0.8745 | N |
| Correlation with Mean Shave at Truth | | | |
| **Metric** | **Corr.** | **$\rho$-value** | **Significant?** |
| $KLnorm$ | 0.3794 | 0.0047 | Y |
| $L_1\,norm$ | 0.1610 | 0.2447 | N |
| $L_2\,norm$ | -0.1001 | 0.4712 | N |
| $L_\infty\,norm$ | -0.1147 | 0.4087 | N |

Table 3: Correlation between metrics evaluated at truth and both efficiency and the amount of shaving, considering all 54 conditions (Significance at 0.05 level)

close to the reference, VCG mechanism, the Small mechanism makes the expected payoff, conditioned on success and failure, both relatively close to the profile under VCG (compared to the other mechanism rules); i.e., close to zero for success and close to forfeiting the maximal payoff for failure. Since the VCG payoff distribution is skewed such that many agents have only small opportunities for gain (see Figure 2), then many of these opportunities can be addressed by the Small mechanism with the remaining opportunities for gain entailing significant risk.

### 4.3 Metric Analysis

In this section we adopt the correlation between each metric and the equilibrium shave factor and efficiency as a measure of the informativeness of the metric in quantifying the degree of strategyproofness of a mechanism. The correlation is determined over a data set of several thousand instances. For each generator (Uniform, Decay and Super) there are 6 mechanisms[8] and 3 different equilibrium analy-ses (for 1, 2 and 3 shave factors.) This provides 3 x 6 x 3 = 54 data points, with the average efficiency, average shave factor, and metric computed for each and enabling a correlation to be computed. The results are presented in Table 3. We only present results for normalized metrics throughout this section because they dominate in terms of statistical significance. *We see that the KL-norm metric is negatively correlated with efficiency and positively correlated with the equilibrium shave factor. In both cases this correlation is significant at the 0.05 level, whereas the correlation for the other, regret-based metrics is not significant.*

Although of secondary importance, we can also consider the informativeness of each metric in *validating* how close to truthful an equilibrium is, based only on observed data in the equilibrium. This is interesting, for example, in evaluating the degree of strategyproofness of a mechanism based only on observed, equilibrium behavior. The correlation data, evaluated over the same 54 conditions but now in equilibrium for each mechanism, is presented in Table 4. We find that the $L_1\,norm$ is more informative, in equilib-rium, than the $KLnorm$ and other metrics. A strong, and significant correlation is also found for the $L_2 norm$ metric. The $L_1 norm$ measures the average (normalized) regret of an agent. Our hypothesis for why the average equilibrium regret is effective in this regard, is that the further a mechanism is from being strategyproof, the further agents will deviate from truthful bidding in equilibrium, and the more mistakes (*ex post*) that will occur. *Note, though, that the $L_1 norm$ metric does not provide guidance for design because it requires a designer to reason about properties in equilibrium. In fact, for a fixed distribution on agent reports (e.g., at truth) almost all of the mechanisms have the same $L_1 norm$ metrics (see Table 1).*

---

[8] We drop Equal, No Discount and VCG from this correlation analysis; No Discount and VCG are not in the candidate class of mechanisms, and Equal is outside the class we are especially interested in because it sometimes allocates an agent more than its VCG payoff.

| Correlation with Efficiency in Equilibrium | | | |
|---|---|---|---|
| **Metric** | **Corr.** | **$\rho$-value** | **Significant?** |
| $KLnorm$ | -0.4989 | 1.2292e-04 | Y |
| $L_1 norm$ | -0.6460 | 1.3269e-07 | Y |
| $L_2 norm$ | -0.5119 | 7.6150e-05 | Y |
| $L_\infty norm$ | -0.3762 | 0.0051 | Y |
| Correlation with Mean Shave in Equilibrium | | | |
| **Metric** | **Corr.** | **$\rho$-value** | **Significant?** |
| $KLnorm$ | 0.2702 | 0.0482 | Y |
| $L_1 norm$ | 0.5870 | 3.0820e-06 | Y |
| $L_2 norm$ | 0.4615 | 4.4464e-04 | Y |
| $L_\infty norm$ | 0.3738 | 0.0054 | Y |

Table 4: Correlation between metrics evaluated at equilibrium and both the efficiency and the amount of shaving, considering all 54 conditions (Significance at 0.05 level)

| Mechanism | $KLnorm$ | $L_1 norm$ | $L_2 norm$ | $L_\infty norm$ |
|---|---|---|---|---|
| Two Triangle | 0.0820 | 0.6096 | **0.3271** | 0.1976 |
| Threshold | 0.0556 | 0.6991 | **0.2984** | **0.1367** |
| Reverse | 0.1421 | 0.9415 | 0.4896 | 0.3104 |
| Small | **0.0452** | **0.5903** | 0.4208 | 0.3534 |
| Large | 0.0668 | 0.8269 | 0.4494 | 0.2916 |
| Fractional | 0.1303 | 1.1456 | 0.5683 | 0.3477 |
| Equal | 0.2033 | 1.3758 | 0.7291 | 0.4919 |
| No Discount | 0.3114 | 1.9962 | 1.0311 | 0.6721 |

Table 5: Metric value at equil. averaged across all three scenarios and equil. classes. Minimal values in **bold**.

In Table 5 we present the various metrics evaluated at the equilibrium of each mechanism over the 54 conditions. Here, it is apparent that Small is most effective at minimizing $L_1 norm$, i.e., in minimizing the average regret faced by agents in equilibrium. In contrast, and counter to accepted wisdom, the Threshold rule (which is designed to minimize maximal regret given reports) has higher average regret in equilibrium. The Threshold rule is most effective in minimizing the $L_2 norm$ and $L_\infty norm$ metrics, which is perhaps unsurprising given its design.

## 5   Online Mechanism Selection

In this section, we adopt a straw-man experiment to understand the effectiveness of the various metrics in guiding an online search for the best mechanism, using only information that is available to an observer in equilibrium play. Note that a simpler question about heuristic design was already answered earlier: the Small mechanism has the best $KLnorm$ metric, and thus would be adopted as the best mechanism design under this lens. But here we ask a different question: given observed equilibrium play, is the $KLnorm$ metric effective in suggesting a new mechanism to switch to? The set-up is one of online search. We do not get to evaluate the counterfactual equilibrium that would exist under each candidate equilibrium, nor the true, underlying efficiency of an equilibrium. The only data that is available is based on observing the equilibrium bids, allocations and payments in a current mechanism.

The online search is instantiated for a particular metric and proceeds as follows. The search takes place over a sequence of epochs, with a single mechanism deployed in each epoch and an epoch consisting of a fixed number of CE instances. The search is initialized somehow (here we always initialize to the No Discount mechanism.) An epoch provides two kinds of data. For the mechanism that is used, it provides distributional information about the equilibrium bids and the metric can be evaluated on the (revealed) payoffs received by agents. But it is also possible to take the same distribution on bids, and evaluate the metric for each of the other available mechanisms. That is, take the bids as fixed and simply evaluate the metric on the payoffs that would be induced by the other mechanisms (and ignoring that the input is actually the equilibrium for the current mechanism, and not the truthful distribution.)

At the end of each epoch, we evaluate each metric based on the data collected in the equilibrium of the current mechanism and switch to the mechanism with the lowest metric. In evaluating the metrics, we retain data from previous runs of the same mechanism as adopted in the current epoch, enabling ever more accurate metrics to be calculated. The only caveat is that we check for cycles and break them as follows: e.g., suppose we are presently using mechanism $A$ and the metric over the data under $A$ indicates mechanism $B$ to be best, but $B$ has been selected in the past and the data under mechanism $B$ indicates that mechanism $A$ is best. If such a cycle is found, then the online search proceeds by evaluating the metric on $A$ and $B$ over the *combined* data set from running both $A$ and $B$ in the past and selecting the best.

Figure 8 shows the results of running this algorithm for each of the three different CE scenarios and for both 1 and 3 agent classes in defining the simulated equilibrium. We compare the performance of the algorithm with the $KLnorm$ and $L_1 norm$ (average regret) metrics. Each graph shows the epoch on the x-axis and the efficiency of the chosen rule as a fraction of the ideal rule (Small) on the y-axis; the epoch size was set to 100 for these experiments. Online search with the $KLnorm$ metric very quickly chooses a good rule, and with performance that tends to dominate that of search with the $L_1 norm$ metric. Performance of the $L_2 norm$ and $L_\infty norm$ is nearly identical to that of $L_1 norm$, and is thus omitted for clarity. The online search performs least well in the Super scenario 3 class case, where it chooses to leave the Small rule for Large based on the data available after epoch 3 and then fails to return. From within its own equilibrium the Large rule looks promising and the ideal Small rule is extremely different in effect and distribution– making escaping the Large local-maxima difficult.

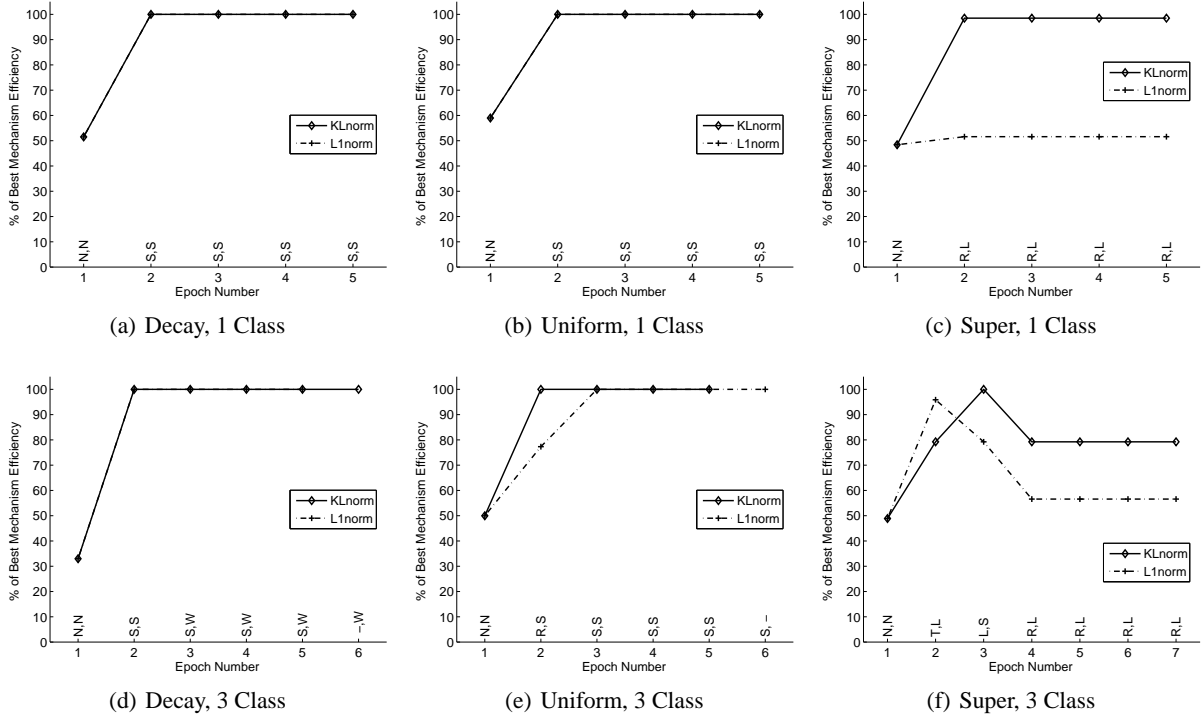

Figure 8: Online selection: choosing the mechanism algorithmically. The labels along the x-axis indicate the rule chosen in a given epoch under the $KLnorm$ and $L_1 norm$ rules respectively, using the abbreviations defined in the Appendix.

## 6    Conclusions and Future Work

The KL-distance metric is defined on the difference between a distribution on agent payoffs in a mechanism and that under a reference, strategyproof mechanism, both evaluated with respect to the true distribution on agent valuations. This metric is shown to be more informative, in terms of correlating with the deviation from truthful bidding in equilibrium, than other regret-based metrics. As a consequence, we also observe that minimizing maximal *ex post* regret (given truthful bids) does not necessarily lead to optimal designs; e.g., the Threshold mechanism is designed this way, but the Small mechanism generates a bet-ter (closer to truthful) equilibrium while also minimizing average regret in equilibrium. In the context of CEs, our results establish that by seeking to match the payoffs in a reference mechanism in distribution, a mechanism designer can achieve a mechanism that is maximally strategyproof in the sense of minimizing the amount by which agents will deviate from truthful bidding in equilibrium.

A number of opportunities exist for future work. It will be interesting to try to directly exploit the $KLnorm$ metric for the *low-level* design of mechanisms in the CE domain; i.e., look to design payment rules that will explicitly seek a distribution on payments (and thus payoffs) that closely approximates that of the VCG mechanism? Second, we can consider different domains for which the VCG mechanism still provides the strategyproof benchmark, such as combinatorial auctions with core constraints or sponsored search with constraints that mandate "simple" payment rules. Third, we would like to consider a mechanism design problem in which the VCG mechanism does not provide the strategyproof benchmark, for example in application to redistribution mechanisms [5]. We can also look to couple the framework with approximation algorithms; i.e., the motivation for the design question here was to circumvent an impossibility result, but what if the motivation was computational intractability? We should elaborate on our hypothesis that alignment with the payoff distributions in a reference mechanism is useful because it selects mechanisms that for a large number of agents provide no advantage to deviation while leaving opportunities for only a small number of agents, and thus a risky strategic proposition. Finally, we propose to directly compute the KL-distance metrics on the multivariate payoff distributions.

## Acknowledgments

Thanks to the anonymous reviewers for their extremely helpful comments on an earlier draft. Portions of this work are supported by a Yahoo! research and KTC grant. David Krych earlier pursued an empirical analysis of this family of CE mechanisms in a Harvard College Undergraduate Thesis, 2003.

## Appendix: CE Mechanisms

The CE mechanisms that we study are all from Parkes et al. [14], except for *Two Triangle* (introduced here):

**(E)qual:** Simply split the available surplus equally among the trading agents.

**(F)ractional** Allocate surplus in proportion to the VCG discounts.

**(S)mall** Allocate surplus from smallest $\Delta_{\mathrm{vcg},i}$ to largest, never exceeding $\Delta_{\mathrm{vcg},i}$.

**(L)arge** Allocate surplus from largest $\Delta_{\mathrm{vcg},i}$ to smallest, never exceeding $\Delta_{\mathrm{vcg},i}$.

**(T)hreshold** Allocate surplus to minimize the maximum $\Delta_{\mathrm{vcg},i} - \Delta_i$, subject to $\Delta_i \le \Delta_{\mathrm{vcg},i}, \forall i \in N$.

**(R)everse** Allocate surplus to maximize the minimum $\Delta_{\mathrm{vcg},i} - \Delta_i$, subject to $\Delta_i \le \Delta_{\mathrm{vcg},i}, \forall i \in N$ (and allocating all of the surplus).

**(W)Two Triangle** Allocate half of the surplus by Threshold and then run Small with the residual.

The **No Discount** mechanism simply has each agent pay its reported valuation for the trade. The Equal mechanism is the only rule in which an agent's discount may be greater than in the VCG mechanism. Each of the mechanisms were designed to minimize different distance metrics between allocated payoffs (or discounts) and VCG payoffs. For example, the Threshold rule minimizes the maximal difference to VCG payoffs across all agents.